\documentclass[twocolumn,showpacs,showkeys,preprintnumbers]{revtex4}
\usepackage{graphicx}
\usepackage{dcolumn}
\usepackage{bm}
\usepackage{color}

\begin{document}

\title{A novel approach for deducing the mass composition of cosmic rays from lateral densities of EAS particles}
\author{Rajat K. Dey}
\email{rkdey2007phy@rediffmail.com}
\author{Sandip Dam}
\email{sandip_dam@rediffmail.com}
\author{Animesh Basak}
\email{animeshbasak75@gmail.com}
\affiliation{Department of Physics, University of North Bengal, Siliguri, WB 734 013 India}

\begin{abstract}
A Monte Carlo (MC) simulation study of cosmic ray (CR) extensive air showers (EAS) has been carried out in the energy regime of the KASCADE experiment. From the characteristics of lateral distributions of electrons and muons of simulated EAS, some important EAS observables are extracted by a novel approach, and their CR mass-sensitivity is demonstrated. The study takes into account the issue of the experimental lateral density profiles of EAS electrons and muons after introducing the notion of the local age and segmented slope parameters, aimed to extract information on CR mass composition from observed data. The estimated lateral shower age and slope from the analysis of the KASCADE data (KCDC) agrees with the idea of a gradual change of CR mass composition from light to heavy around the knee.
\end{abstract}

\maketitle

\section{Introduction}

Cosmic rays (CRs) of galactic and extragalactic origins are the ultra-high-energy (UHE) particles which stream into our atmosphere from distant corners. Ground based detector arrays are designed for the measurement of their composition and energy spectrum by detecting the  secondary particles of the extensive air showers (EAS) induced by the CR particles. Electrons ($\rm{e}^{-}+\rm{e}^{+}$) and muons (${\mu}^{-}+{\mu}^{+}$) in the EAS constitute the major contribution to the ground detector signals. Reconstructed densities of these particles at various detector locations, as well as the EAS core, and other EAS parameters are then obtained from the analysis of detector signals or crude densities of particles. In an EAS experiment, the electron and muon densities, and their arrival times, are known to be the chief observables to extract any information of the primary cosmic rays (PCRs).

The modeling of lateral density distribution (LDD) of EAS particles through Monte Carlo (MC) simulations is needed in any experiment in order to interpret the data. By applying the shower reconstruction to the simulated and observed LDD data of electrons independently exploiting a suitable lateral density function (LDF), the important shower parameters namely the shower/electron size ($N_{\rm e}$), the EAS core and the shower age ($s$) are usually obtained. Some earlier observations (\emph{e.g.} in [1]) and recent simulation studies indicate that the LDD of shower electrons could not be described properly with a single age at all distances, which indicates that the lateral age varies with the core distance. This radial dependency on the lateral shower age by the NKG function could not be nullified totally from the several subsequent modified forms of the structure function. At this juncture, the notion of variable lateral age or local shower age parameter (LAP; ${\rm{s}_{\rm{local}}^{\rm{electron}}(\rm{r}})$) of an EAS was emerged [2,3]. In recent past, we have observed that the nature of variation of the local age with radial distance maintains nearly the same configuration independent of the primary CR energies. The above study indicates that the local age as a function of the radial distance from the EAS core exhibits some sort of scaling nature. 

In this work, we have extended the concept of the scaling nature of LDD of electrons in terms of the LAP defined in earlier works, to LDD of muons by exploiting simulation data from a MC CR shower generator {\emph{CORSIKA}} [4]. In order to substantiate the predictions by the simulation, we also examine the possibility of utilizing experimental LDD data of electrons and muons, freely available from KCDC data shop [5] of the KASCADE experiment for the purpose of comparison and interpretation. 

 The NKG type LDF is a solution arising out of the implementation of several approximations applied to the 3D diffusion equations of an EM cascade in the atmosphere. It is indeed a limiting LDF of EAS electrons through the saddle point solutions in approximation A and zero ionization losses of electrons in approximation B. The NKG type LDF, originally derived for the electromagnetic (EM) cascade was modified in two ways for the purpose of describing observed or simulated muon densities for hadron initiated showers to obtain shower parameters. In the first approach, the scale distance or the Moliere radius ($R$) was adjusted in the LDF (values are usually chosen as 320 and 420 m in the KASCADE data analysis). This LDF is actually used for estimating LAP for muon distributions in this work. Secondly, this LDF is further modified by adjusting its slope or exponent in ($\frac{r}{R}$) and ($1+\frac{r}{R}$) functions for estimating the segmented slope parameter (SSP; ${\beta_{\rm{SS}}^{\rm{muon}}(\rm{r}})$). The overall LDF with an arbitrary exponent ($\beta$), is used as Greisen-like LDF with Greisen radius ($R=R_{G}$) equal to either $320$ or $420$ m in the analysis. Just like the LAP, the SSP is in essence the slope of the muon LDF at a point. It is clear that slope of a LDF is directly linked with the shower age parameter. This hints that LAP and SSP are two different manifestations of the same physical information, only difference in their mathematical depictions. For a multi-observable study of EAS the SSP and LAP have been used to correlate the slope/age with other EAS observables to identify the nature of the shower initiating primaries. We have taken truncated muon sizes in the simulation for a better comparison with the KASCADE data by counting muons in the core distance interval $40\textendash 200$ m. 

In this paper we address some aspects of the LAP and SSP arising out of the LDD of EAS electrons and muons, worked out by a detailed simulation study. We check the validity of the scaling behavior of LDD of muons through the LAP and also the SSP as hinted by our earlier works [2]. Our main focus is to examine the possibility of utilizing the LAPs and SSPs estimated from the available KCDC data, in order to deduce CR mass composition.

The plan of this paper is the following. In section 2 we recall the shape parameters describing the lateral developments of EAS electrons and muons, and definition of the LAP and SSP, is given. In section 3 the CORSIKA events used in the procedure of local shape parameter estimation are described. Section 4 is devoted to the data analysis for the estimation of the LAP and SSP. Our results on the primary CR mass sensitivity of these EAS shape parameters are presented and discussed in section 5. Section 6 summarizes our final conclusions.           

\section{Local age and segmented slope parameters}

The shape parameters \emph{viz.} the shower age/ slope,  describing the lateral distributions of the EM component of a shower first came into existence in CR studies through the works of Moli\'ere [6], Rossi and Greisen [7], and lately by Nishimura [8]. During 1956-60, Kamata and Nishimura, and then Greisen independently demonstrated [9] that for hadron initiated EAS, the longitudinal and lateral profiles of the EM component can be described by some average overall single shower, assigning some suitable values to the shape parameters.      
 
There has been a numerous number of articles and reviews written by many authors on the concept of shower age or slope and their estimations, for several decades [3,10-13 and references therein]. A more judicious estimation of lateral shower age parameter has been made through LAP in the work [2,13]. In this work we have made an attempt to estimate the both; LAP and SSP from LDDs of electrons and muons, and to derive possible conclusions on the nature of EAS  generating primary CR particles.   

Observed data of many EAS experiments [13] validated the theoretical prediction by Capdevielle [3] with regard to the erroneous estimation of shower size and the shape parameter using the NKG type LDF. He then resolved the problem in bringing the idea of local shape parameter in place, for the better description of LDD of electrons [2,3]. 

According to the concept of local shape parameter, we can now give its analytical expression for an arbitrary LDF in NKG formalism (say, $\rm{f}(\rm{x})$ with $~\rm{x}={\rm{r}\over \rm{R}_{\rm{D}}}$) between two adjacent points [$\rm{x_{i}},\rm{x_{j}}$]~: 

\begin{equation}
a_{\rm{local}}(i,j) = {{\ln(F_{ij} X_{ij}^{\alpha_{1}} Y_{ij}^{\alpha_{2}})} \over {\ln(X_{ij} Y_{ij})}}
\end{equation}

The following substitutions were made in obtaining the local shape parameter: $F_{ij}$ = {{$f(r_{i}$)}/{$f(r_{j}$)}}, $X_{ij}$=$r_{i}$/$r_{j}$, and $Y_{ij}$=($x_{i}$+1)/($x_{j}$+1). More generally, if $r_{i} \rightarrow r_{j}$, this suggests the definition of the shape parameter $a_{local}(x)$ (or $a_{local}(r)$) at each point~:

\begin{equation}
a_{\rm{local}}(x) = {1 \over {2x+1}} \left( (x+1) {{\partial{\ln f}} \over {\partial{\ln x}}} + (\alpha_{1}+\alpha_{2})x + \alpha_{1} \right)
\end{equation}

The identification $a_{local}(r)\equiv a_{local}(i,j)$ for $r=\frac{r_{i} +r_{j}}{2}$ remains valid for the experimental distributions (taking $F_{ij}~ =~\rho(r_{i})/\rho(r_{j}$)$\equiv ~f(r_{i})/f(r_{j}$)).

In cascade theory, the solution of the 3D diffusion equations under some approximations can provide the LDF of cascade particles by the
well-known NKG structure function [14], given by

\begin{equation}
{\rm f}(r)={\rm C}(s_{\perp})(r/R_{\rm m})^{s_{\perp}-2}(1+r/R_{\rm m})^{s_{\perp}-4.5} \;,
\end{equation}
where the normalization factor C($s_{\perp}$) is given by 
\begin{equation}
{\rm C}(s_{\perp}) = \frac{\Gamma(4.5-s_{\perp})}{2\pi\Gamma(s_{\perp}) \Gamma(4.5-2s_{\perp})}  \;.
\end{equation}

With the help of the structure function ${\rm f}(r)$ the electron or muon density ($\rho_{\rm {e}/\mu}$) for a constant shape parameter $\rm{s_{\perp}}$ can be given by
\begin{equation}
\rho_{\rm {e}/\mu}(r) = {N_{\rm {e}/\mu} \over R_{\rm m}^2 }{\rm f}(r) \;.
\end{equation}

At the KASCADE level, $\rm{R}_{\rm m}$ took $89$ m and $320$ or $420$ m for the analysis of electron and muon LDD data respectively.  

The Greisen type LDF, used for muons, was obtained from the modifications of the Moli\'ere radius and the slope or exponent of functions $(\frac{r}{\rm{R}_{\rm m}}$) and $(1+\frac{r}{\rm{R}_{\rm m}}$) in the NKG structure function, given by  

\begin{equation}
f_{\mu}(r)={\rm{A}}(r/R_{G})^{-{\beta}}(1+r/R_{G})^{-2.5},
\end{equation}

with the Greisen radius $\rm{R_{G}} = 320$ or $420$ m which was amended in various measurements.

It is clear that for NKG-type LDF, the exponents $\alpha_{1}$ and $\alpha_{2}$ in the relation (1) take values as $4.5$ and $2$. But for Greisen-type LDF, $\alpha_{1}$ and $\alpha_{2}$ would take $0$ and $2.5$ respectively. Moreover, we have modified the denominator in the relation (1) by $X_{ji} = X_{ij}^{-1}$ and $Y_{ij} = 1$ while Greisen-type LDF is being considered. 

 For NKG-type LDF with different $R_{\rm m}$, the $a_{\rm{local}}(r)$ is the so called  LAP, is already denoted by $s_{\rm{local}}(r)$ [13]. Similarly for Greisen-type LDF i.e. $f_{\rm{Greisen}}(r)$, we denote the shape parameter by $\beta_{\rm{ss}}(r)$, is called the SSP.

For the observed LDD data of electrons, the characteristics of LAP as predicted in [15] were reaffirmed by the Akeno [16], KASCADE [3], North Bengal University (NBU) [17] and other experiments [3,18]. This work gives emphasis on the estimation of the LAP and the SSP for the LDDs of electrons and muons. The general equation (1) for estimating the local shape parameters (LAP and SSP), contains only the density and the distance data of an EAS event. However, in reference to a real EAS array, these data significantly depend on the accuracy of the density and EAS core measurements, and also the array triggering conditions. Hence, LAP and SSP have to be estimated within some selective physical intervals of distance, keeping in mind the degree of shower to shower fluctuations, array size, array resolution and intermediate distances among detectors. It would be an interesting task to apply the present method to the LDD data of electrons and muons obtained from GRAPES - 3 experiment at Ooty [19]. The experiment contains a densely packed air shower array and presently a world largest size muon detector ($\approx{1120}~\rm{m^{2}}$) for the measurement of CR mass composition. We have taken selective intervals between two successive points on the radial distance scale with increasing distance keeping, $\Delta_{ij}=\ln{r_{j}} - \ln{r_{i}} \approx{0.5}$ on the log scale.

\section{The Monte Carlo data sample}

A sample of EAS events has been generated by using CORSIKA program version 6.970/7.400 [4]. The simulated events have been generated by using the EPOS 1.99 [20] interaction model for the high-energy hadronic interactions, in combination with the UrQMD model [21] for the low-energy hadronic interactions. Two smaller samples of the MC events have also been generated with combinations, QGSJet01.c [22] - UrQMD and EPOS-LHC [23]-UrQMD. The EM interactions of the EAS is implemented by means of the EGS4 [24] program package. Only proton and iron showers have been generated for obtaining an indication on the probable mass composition of PCRs from KASCADE data. The simulated events in the primary energy range $10^{14}$ eV to $3 \times 10^{15}$ eV have been generated according to the following distribution, given by

\begin{equation}
N(E)dE =N_{0}E^{-\gamma}dE.
\end{equation}
 
The spectral index in the above energy range has been taken as $\gamma = 2.7$ and a total of about $2.5\times 10^5$ showers have been generated. Beyond this energy range, about $5000$ events have been generated with $\gamma = 3.1$. Showers have been generated at the KASCADE level by setting kinetic energy cut-offs for different secondary particles in the CORSIKA steering file, restricting  zenith angle ($\Theta$) in the range $0^{o} - 18^{o}$ as per requirements of the KASCADE experiment.    

\section{Shower data analysis: Estimation of the LAP and SSP}

We have computed the LAP for the LDD of electrons and muons for each simulated and observed event independently by using equation (1). The NKG-type LDF was used for ${\rm f}(r)$ in the equation. The Moli\'ere radius was taken as $89$ m for electrons while for muons, $R_{\rm{m}}$ took values as $320$ m and $420$ m in two different situations, as used in the KASCADE data analysis. For computing the SSP, the so-called Greisen LDF was chosen with the Greisen radii $320$ m and $420$ m independently. The estimation of the LAP from simulated LDD data of electrons is affected by an average statistical error of the order of $\pm{4 - 5}\%$ for $7~\rm{m} < \rm{r} \le 300$ m around the knee energies. However, for $\rm{r} < 7$ m and $\rm{r} > 300$ m the error for the LAP may rise up to about $\pm{10}\%$. The LAP and SSP estimated from simulated muon data receive about $\pm{1 - 2}\%$ higher fluctuations for $7~\rm{m} < \rm{r} \le 300$ m in the concerned energy region. The sources of statistical error are due to mainly in the fluctuations in electron/muon densities and the uncertainties in core distance measurement. Simulated data provide very accurate position of each EAS particle, and hence the core distance estimation contributes reasonably insignificant errors to the estimations of the LAP and SSP. For controlling the statistical fluctuations over the estimations in  electron/muon densities at different distance bands, a reasonable sample size of EAS events should be ensured.

The KASCADE data from KCDC are available in the form of deposited electron and muon energies in the unshielded and shielded scintillation detectors. We have found these data from KCDC, distributed in various detector positions from near $10$ m to far $200$ m and above core distances. The muon LDD data in the KASCADE experiment were fitted in the $40 - 200$ m radial distance range to obtain EAS parameters. These fitted parameters are available in a different root file in KCDC. It is noticed that the quality of muon data given in KCDC root file are very unreliable above $\approx{120}$ m range for estimating the LAP and SSP systematically. Accordingly we have restricted our analysis with observed muon data in the core distance range about $10 \le{\rm r} \le 120$ m. The main observables in the work are mean minimum LAP and mean maximum SSP. These parameters have been effectively estimated at about 44 m (LAP) and 71 m (SSP), that lie above 40 m range. It can be argued that muon densities below 40 m do not have any serious impact on the estimation of these observables. The deposited energy data obtained from detectors have been converted to electron and muon numbers and then to their densities in the analysis. For converting energy data into densities, we have used their so-called lateral energy correction functions (LECFs) for  electrons and muons available in the KASCADE report and references therein [25].  It is noticed that the LAP estimated from KASCADE electron data experiencing little higher fluctuations, compared to simulation results. For observed muon data, fluctuations in the measurement even cross these limits, as contributed by the observed electron LDD data. 

The variation of the LAP and SSP with radial distance from the EAS core is a fundamental study of this paper. This is the fist time, the LDD of muons have been analyzed for studying the radial dependence of the LAP/SSP, thereby exploring the primary mass sensitivity of these parameters with the KASCADE muon data. 

The variations of $s_{\rm{local}}(r)$ and $\beta_{\rm{ss}}(r)$ with $r$ around the knee energy and beyond are shown in Fig. 1 and Fig. 2. Figure - 1(a) represents $s_{\rm{local}}(r)$ versus  $r$ variation obtained from the simulated LDD of electrons for different electron size intervals. To examine whether the experimental data also demonstrates a high-low-high kind of nature in the LAP, we have included the local ages from the LDD data of electrons of the KASCADE experiment for a particular electron size interval and compared these values with our simulation results in fig. 1(b). The experimental data clearly support the trend predicted by the simulation results at intermediate radial distances.\\
 
We study similar radial variations of the LAP for simulated and KASCADE muon LDD data, which are shown in fig. 1(c) and (d). Interestingly, the shapes of the radial variation versus local age curves estimated from the simulated LDD data exhibit almost the same configuration. It is noticed that the LAP estimated from the KASCADE muon data could not follow the simulated results exactly, irrespective of high-energy hadronic interaction models. Furthermore, some previous studies using the LDD of electrons [2] observed that the shape of these variations are independent of the energy of the EAS initiating particles and also the observation level. Though all figures are not shown, but we have noticed, that the above features of the local age are true for muons also.     

On the other hand, in fig. 2(a) and (b), such type of studies are depicted for the parameter $\beta_{\rm{ss}}(r)$ from simulation and KCDC data for muon densities. Here, it appears that the KASCADE results follow the simulation predictions convincingly from both the high energy hadronic models. In fig. 2(a) and (b), we have noticed an opposite trend in the variation of the $\beta_{\rm{ss}}(r)$ parameter relative to $s_{\rm{local}}(r)$ versus $r$. It can be concluded that the LAP and the SSP of electrons and muons exhibit some sort of scaling behavior while varied as a function of the core distance. It should be mentioned that the variations of $s_{\rm{local}}(r)$ or $\beta_{\rm{ss}}(r)$ with $r$ for muons (fig. 1(c), (d) and fig. 2(a), (b)) are reported first time in this work. 
\begin{figure}
\centering
\includegraphics[width=0.4\textwidth,clip]{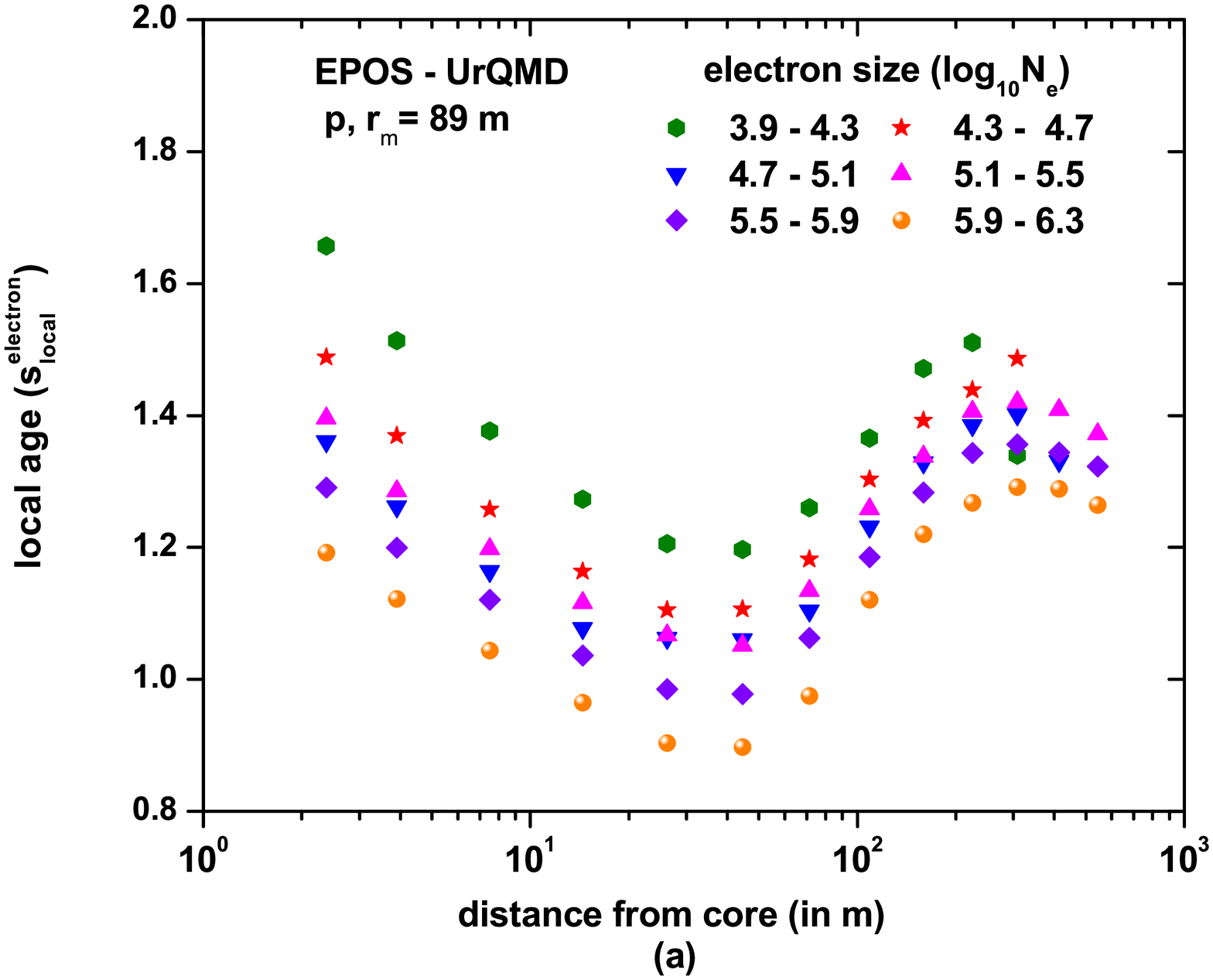} \hfill 
\includegraphics[width=0.4\textwidth,clip]{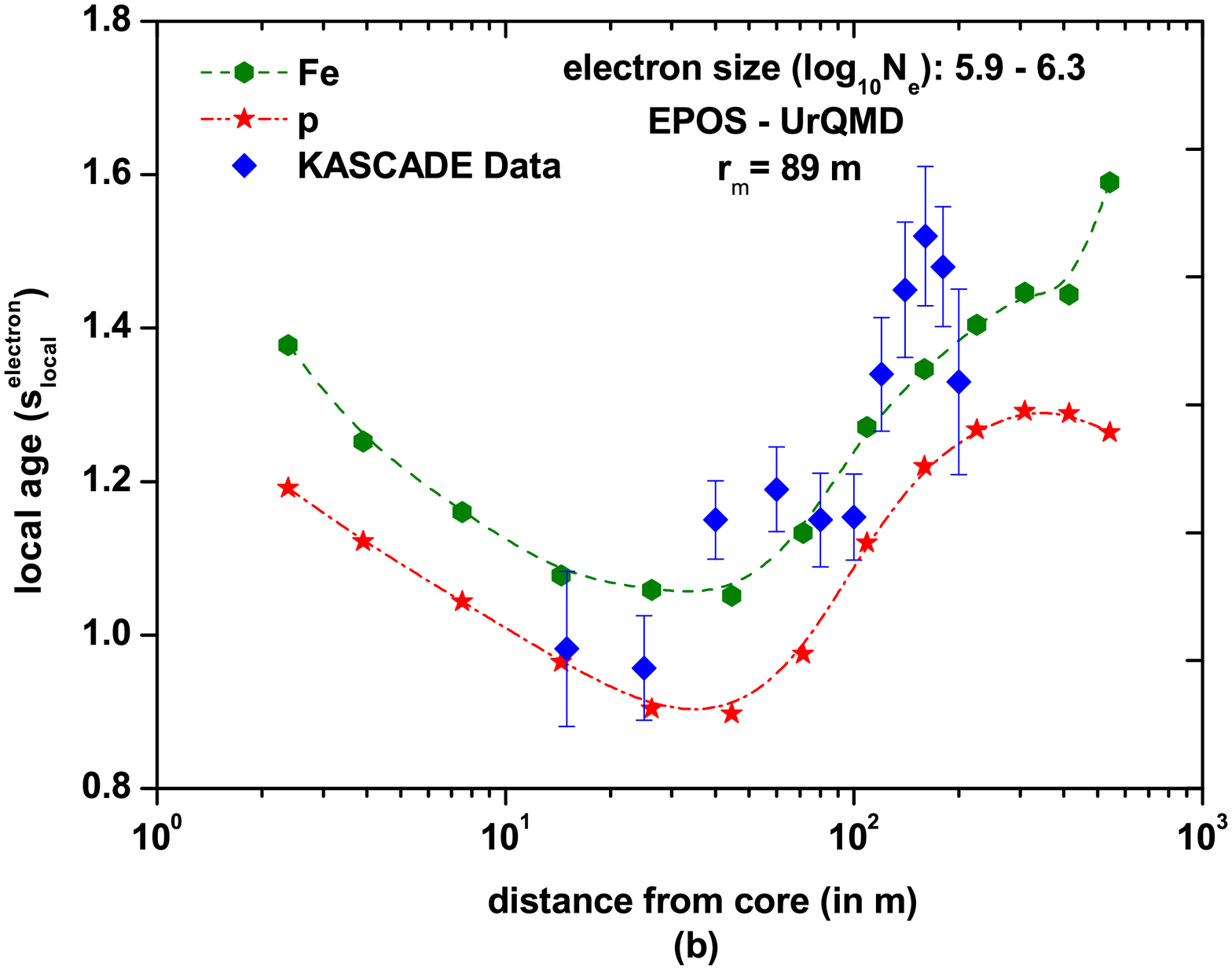} \hfill
\includegraphics[width=0.4\textwidth,clip]{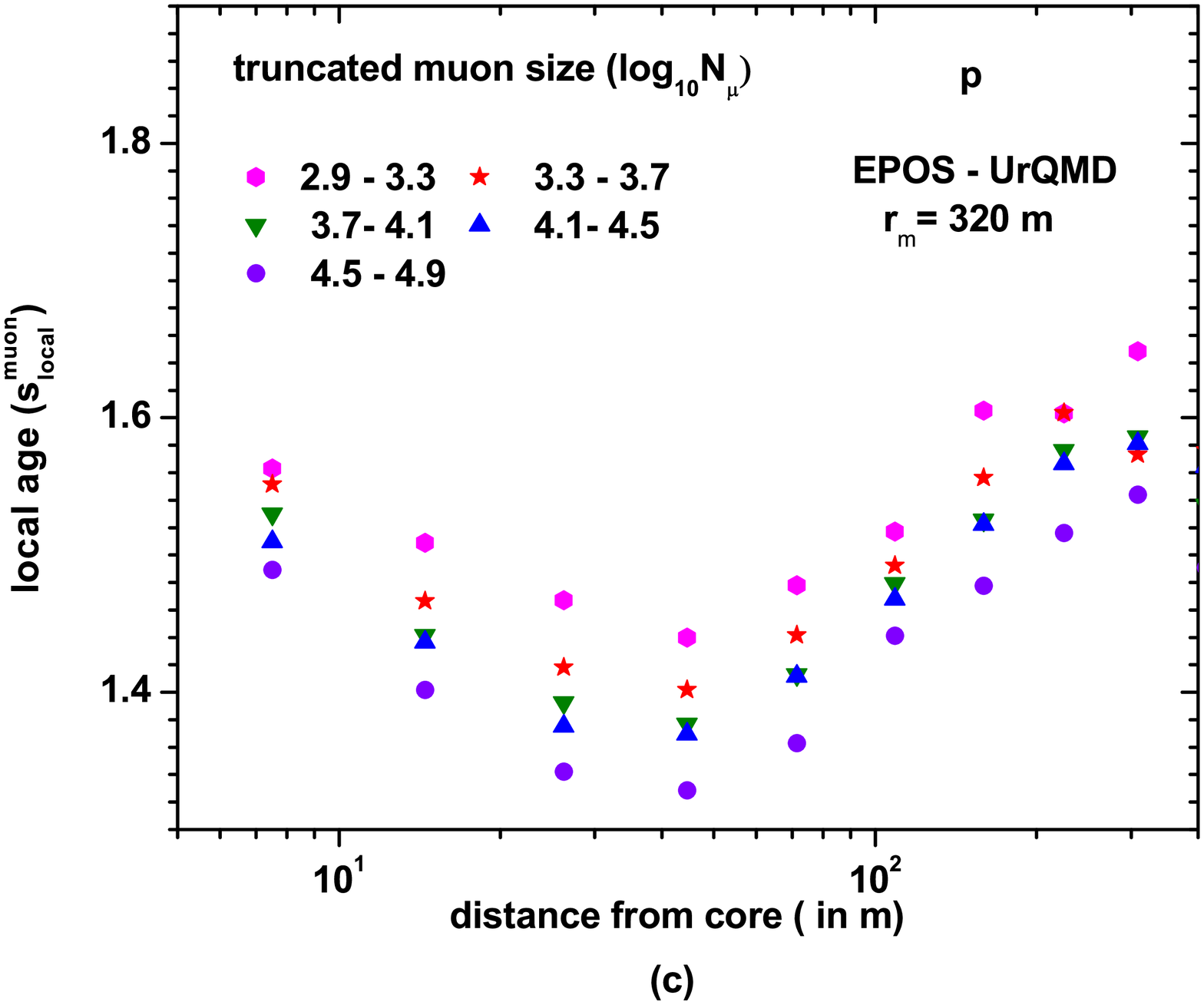} \hfill
\includegraphics[width=0.4\textwidth,clip]{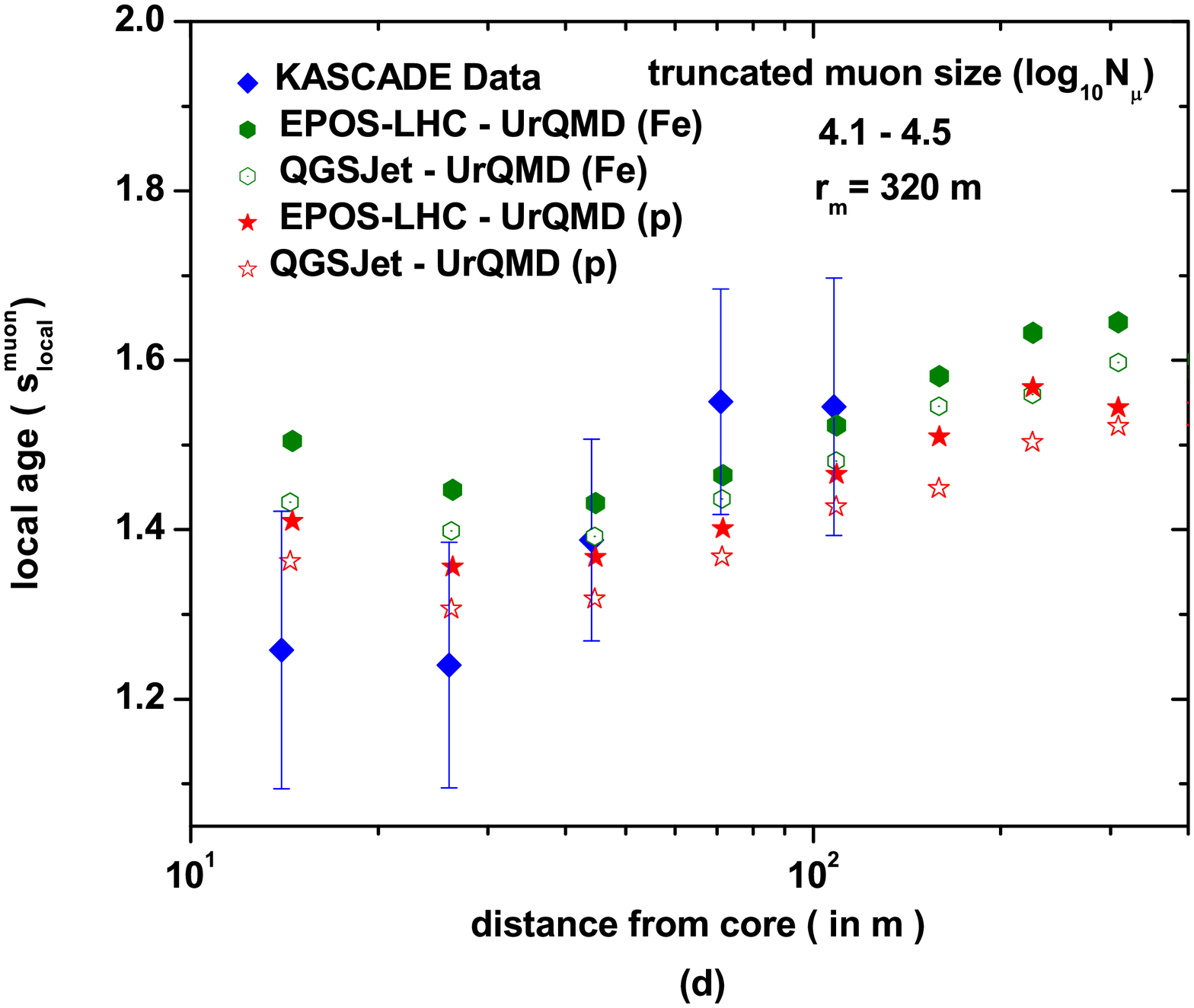} \hfill
\caption{Variation of the LAP (estimated from simulated electron data) with radial distance for different shower (or electron) sizes at the KASCADE site  for (a) p; (b) for both p and Fe along with the LAP obtained from the experimental data. Similar studies are shown for the radial variation of the LAP (estimated from muon LDD data) for different truncated muon sizes in Fig. c and Fig. d. Effect of the high-energy interaction hadronic model on the LAP is shown in Fig. d.}
\end{figure} 
    
\begin{figure}
\centering
\includegraphics[width=0.4\textwidth,clip]{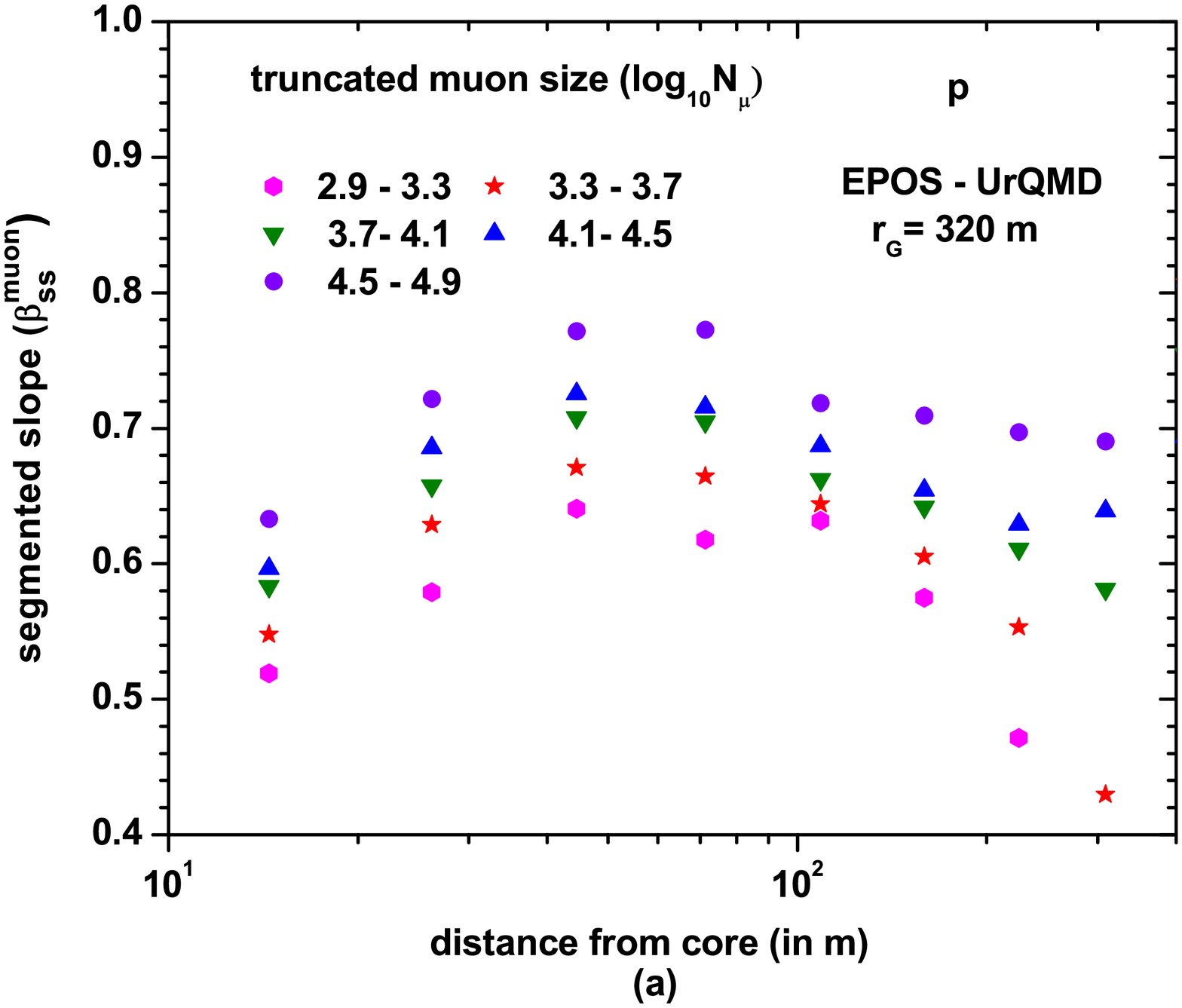} \hfill 
\includegraphics[width=0.4\textwidth,clip]{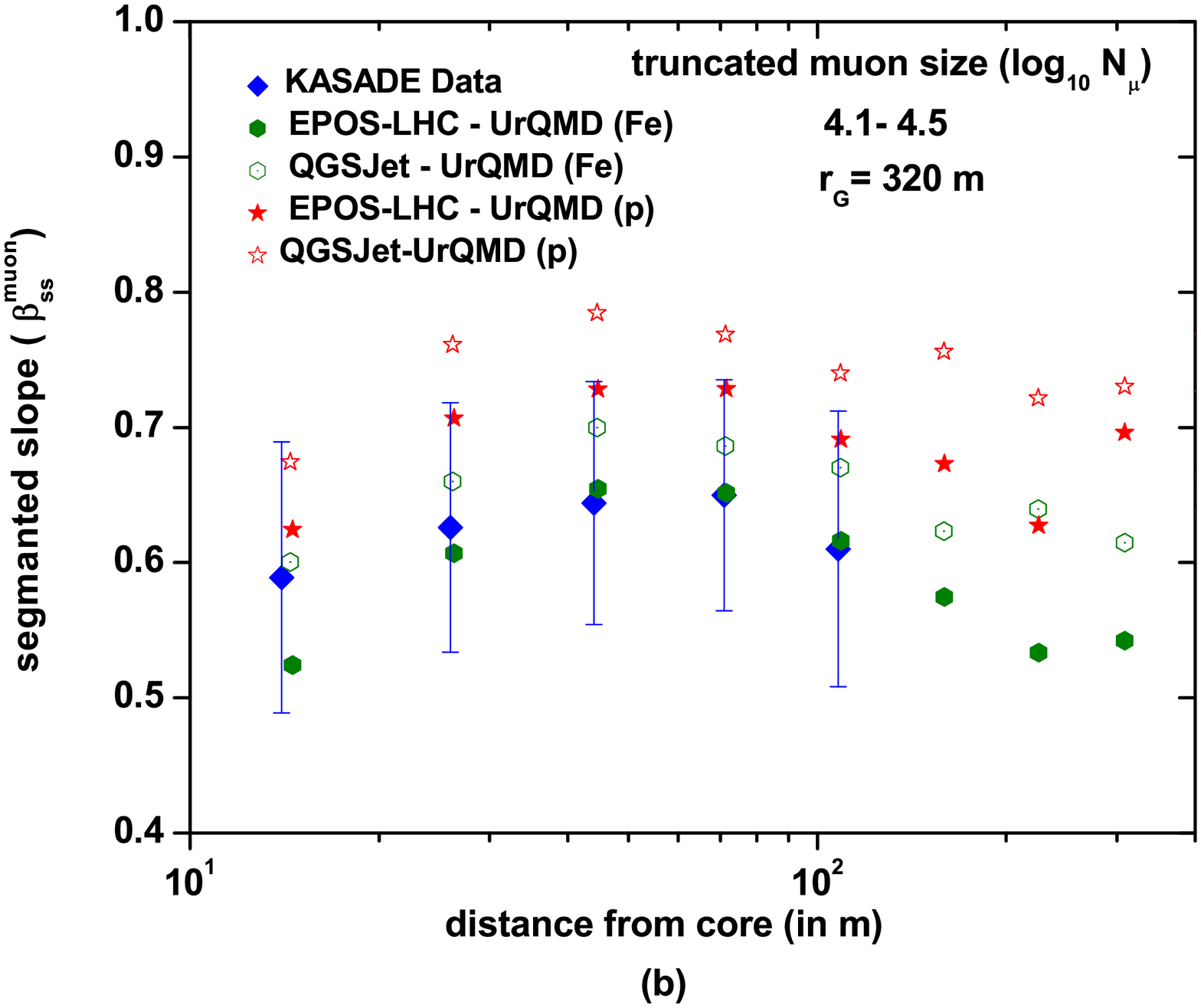} \hfill
\caption{Variation of the SSP (estimated from simulated muon data) with radial distance for different truncated muon sizes at the KASCADE site for (a) p; (b) for both p and Fe along with the SSP obtained from the experimental data. Effect of the high-energy interaction hadronic model on the SSP is shown in Fig. b.}
\end{figure}    

\begin{figure}
\centering
\includegraphics[width=0.4\textwidth,clip]{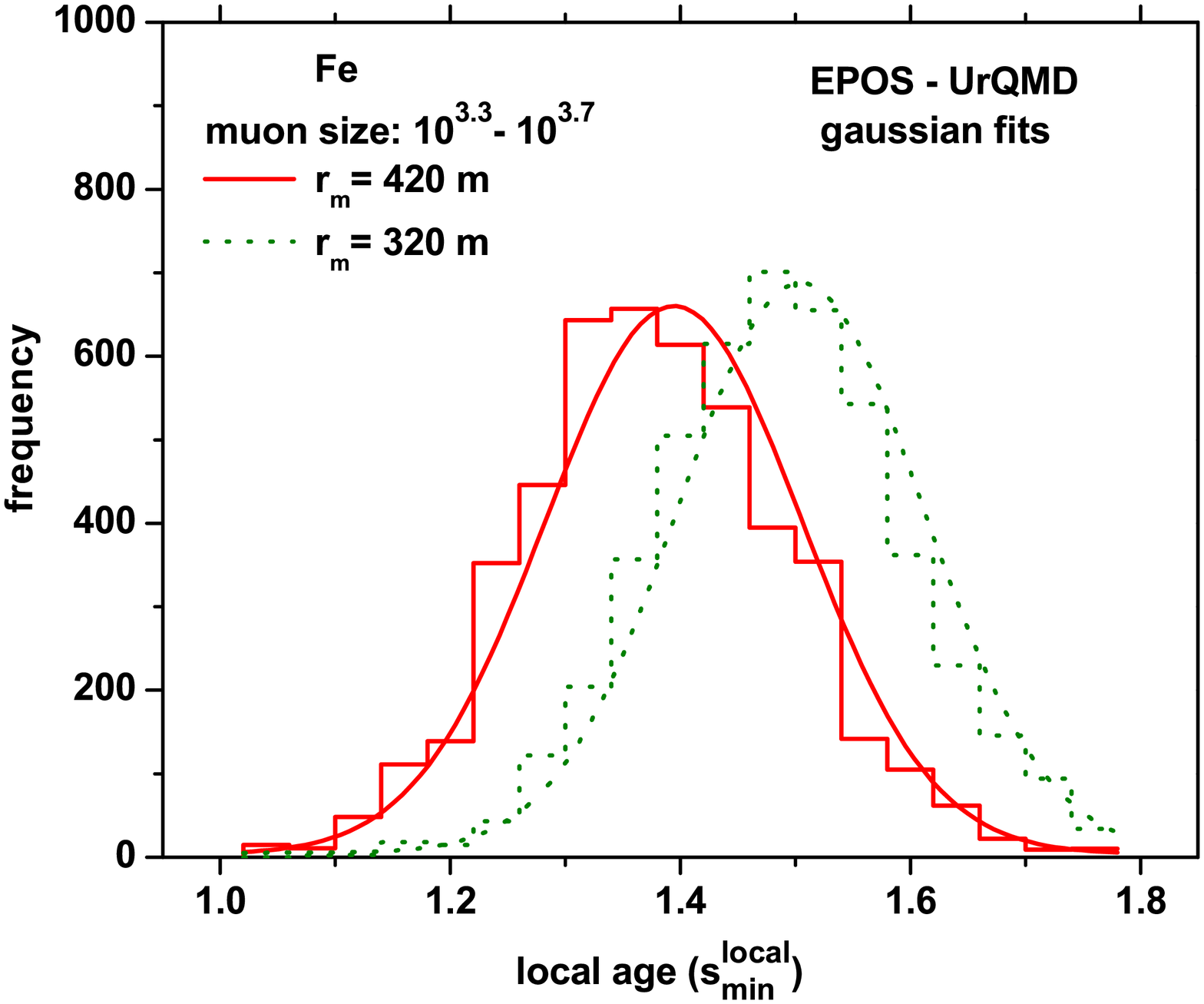} \hfill 
\caption{Distribution of the mean minimum LAP from simulated muon data for iron showers for two different Moli\'ere radii.}
\end{figure}  

If attempts are made to explore the possible causes for such a radial behavior of the $s_{\rm{local}}(r)$ or $\beta_{\rm{ss}}(r)$ or the LDD of electrons/muons, we should at least look at the EGS4 and NKG options closely. In simulations, particularly at the last stage of the cascade development, the generation and absorption of the electromagnetic components are the outcome of different processes/mechanisms implemented in EGS4 package. On the other hand, the LDD data of electrons/muons resulted from the several mechanisms (Moller, Bhaba sattering, positron annihilation etc.) in EGS4 option, were used to estimate the $s_{\rm{local}}(r)$ or $\beta_{\rm{ss}}(r)$ parameters. These parameters were defined in terms of the NKG or Greisen functions (equation ($1$)), that could be obtained from solutions of the cascade equations by applying several approximations. From some previous studies [3,18], it was learned that the behavior of $s_{\rm{local}}(r)$ arises due to the description of EGS4 or observed data by the approximate LDFs (\emph{e.g.} NKG), called shower reconstruction, for obtaining the required EAS parameters.  

The scale distance such as the Moli\'ere or Greisen radius used in the LDF functions may lead to determine the shape of the $s_{\rm{local}}(r)$ or $\beta_{\rm{ss}}(r)$ versus $r$ curves [2]. The above idea of $\rm{R}_{\rm{m}}$ dependent $s_{\rm{local}}(r)$ has been corroborated from the frequency distribution of the minimum values of $s_{\rm{local}}(r)$ (obtained from $r$ versus $s_{\rm{local}}(r)$ plots for two different Moli\'ere radii), which is shown in fig. 3.          

\section{Results concerning cosmic ray composition}

The present analysis is based on the shape of the simulated/observed LDF derived from electrons/muons LDD data. To probe the CR mass sensitivity associated by the shape parameter of electron and muon LDFs, we study some important characteristics of the LAP and the SSP. A shower event is generally characterized by a number of shower parameters.  We are interested to explore the physical nature of the lateral shower age and the slope parameters. Instead of considering several local age or segmented slope values associated with the lateral age or slope, a more systematically chosen/estimated single lateral shower age or slope are expected to be more advantageous to explore the nature of the EAS initiating primary particle here. The minimum value of the local age measured at the radial distance about $44$ m appears to be very effective. Now taking simple average of all these minimum age values of all selected EAS events corresponding to a shower size interval, is defined as the mean minimum local age. The maximum value of the SSP at the radial distance about $71$ m also exhibits its sensitivity to explore the physical nature associated with the slope parameter. It is worthwhile to mention that the low-high-low kind of radial character of the SSP has not been investigated before. The mean maximum SSP can also be obtained from a group of EAS events falling within a muon size interval by applying the above procedure. It is imperative to recall that a high-low kind of radial nature of the local age $\ge{300}$ m from the Akeno and KASCADE-Grande observed electron LDD data was reported in [26]. In terms of the ratio of the measured to the NKG fitted electron densities, a maximal deficit was found at ${50 - 80}$ m core distance range by the KASCADE report [27], and also an excess at large distances.        

\subsection{Variation of the minimum local age with shower size and muon size}

We have plotted the mean minimum local age against mean shower size, obtained from simulated electron LDD data for proton and iron initiated showers at the KASCADE level in fig. 4. In order to judge the effect of the high-energy hadronic interaction model on the above result, if any, we have plotted the above variation for models EPOS 1.99 and QGSJet in the same figure. It is found that the EPOS model yields little higher values for the minimum local age parameter. Results of the KASCADE experiment obtained from the analysis of KCDC electron data are also presented in the same figure. These two plots strongly favor the idea that KASCADE data indicate a gradual change in the CR mass composition across the knee from a lighter towards a heavier, as the shower size/primary energy increases. Using the $N_{\rm e}$-$N_{\mu}$ variations the KASCADE experiment also arrived at the similar inference on the CR mass composition across the knee. We have found a little better results with the interaction model EPOS 1.99 over the QGSJet here.  

\begin{figure}
\centering
\includegraphics[width=0.4\textwidth,clip]{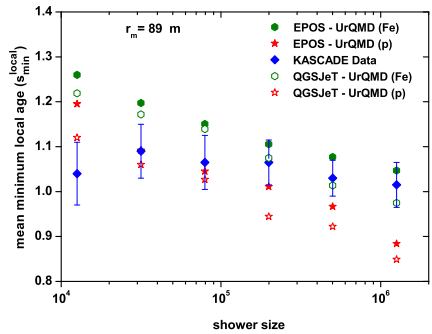} \hfill 
\caption{Variation of the mean minimum local age (estimated from electron LDD data) with shower size for simulated proton and iron primaries for EPOS 1.99 and QGSJet models at the KASCADE site along with the experimental data.}
\end{figure} 

\begin{figure}
\centering
\includegraphics[width=0.4\textwidth,clip]{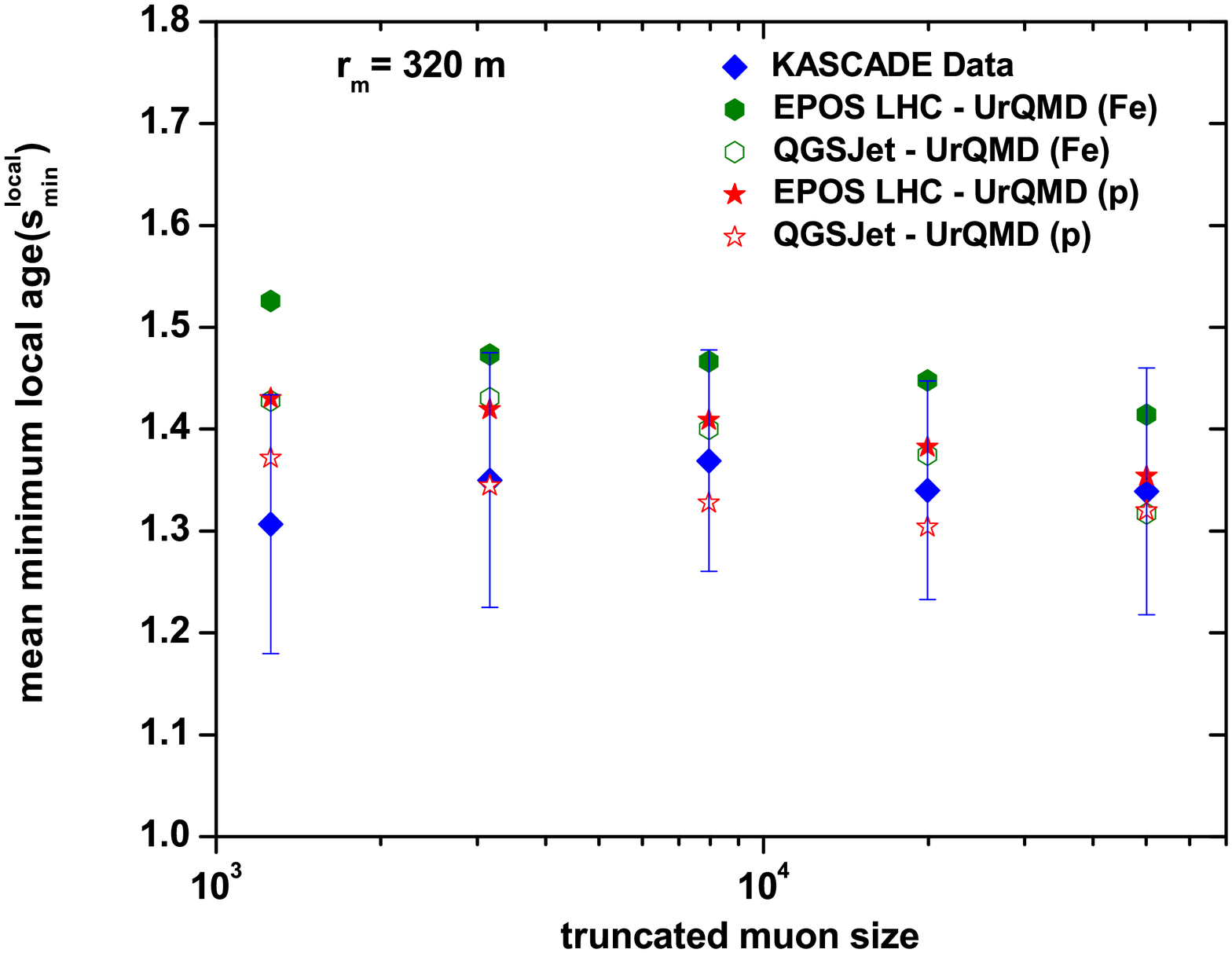} \hfill 
\caption{Variation of the mean minimum local age (estimated from muon LDD data) with truncated muon size for simulated proton and iron primaries for EPOS-LHC and QGSJet models at the KASCADE site along with the experimental data.}
\end{figure}

The variations of the mean minimum local age with muon size, obtained from simulated muon LDD data for proton and iron initiated showers, using the EPOS-LHC and QGSJeT interaction models are shown in fig. 5. The corresponding observed results have been included into the plot, estimated from the KASCADE muon LDD data. It has been noticed here that the mean minimum local age decreases slowly with muon size compared to similar curves found in fig. 4. This might be due to a relatively flatter density profile for muons compared to the case for electrons, which is a generic feature of an EAS. It is revealed that the KASCADE muon data also indicates a heavier domination with increasing muon size/energy across the knee. Small variations are seen in simulated results corresponding to the two adopted high energy hadronic interaction models. The idea of a gradual transition from lighter to heavier composition is somewhat at least not obvious in experimental results.      

\subsection{Variation of the maximum segmented slope with muon size}

\begin{figure}
\centering
\includegraphics[width=0.4\textwidth,clip]{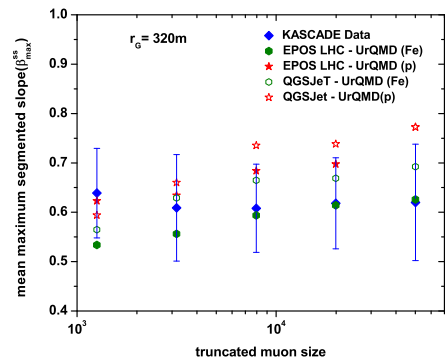} \hfill 
\caption{Variation of the mean maximum segmented slope (estimated from muon LDD data) with truncated muon size for simulated proton and iron primaries for EPOS-LHC and QGSJet models at the KASCADE site along with the experimental data.}
\end{figure}

The variation of the mean maximum segmented slope with muon size for simulated proton and iron showers are shown in fig. 6. The corresponding KASCADE result from the analysis of available KCDC muon data is also shown in the same figure. Two high energy hadronic models are used to interpret observed results. We have seen a reasonable hadronic interaction model dependence of these studies. The observed results for the variation of the mean maximum segmented slope with muon size favors the idea of a gradual transition from lighter to heavier composition of PCRs. All these curves in fig. 6 follow a completely opposite trend compared to the variations found in fig. 4 and 5.

In the present analysis, the errors in estimating the mean minimum local age in the shower size interval $10^4 - 10^6$ result in the range $\pm({0.03 - 0.05})$, irrespective of the high-energy hadronic interaction model.  These errors are comparable with those obtained from KASCADE electron LDD data. In case of observed muon data, errors in estimating the mean minimum local age, as well the mean maximum segmented slope, are found often in the limit $\approx \pm({0.10 - 0.15}$) in the concerned truncated muon size range.
  
\section{Conclusions}
In the present analysis, we have particularly focussed on the LAP and SSP, drawing the following conclusions.

1. The LDD of electrons in a shower manifests some sort of scaling (shower size independent) nature in terms of the LAP. This feature of the LAP still persists when the parameter has been estimated from the LDD of muons. The radial variation of the LAP follows a configuration where with an increasing of the core distance, the parameter decreases initially and attains a minimum, at about $44$ m, then it moves up, attaining a local maximum at about $\approx{300}$ m, and then moves towards a minimum value again. This feature does not change considerably with the shower size and muon size of the EAS. These characteristic variations in the LAP appear as a basic feature of the EM and muonic cascades of a shower.

2. The LDD of muons in a shower also exhibits scaling nature in terms of the SSP. The characteristic feature of SSP versus the core distance curves satisfy a complete opposite configuration compared to the radial variation of the LAP. The SSP estimated from the LDD of muons follows a low-high-low kind of radial variation at least within the range $10 - 300$ m. Here, the SSP attains its maximum value at the core distance $\approx 71$ m.

3. The scale distance such as the Moli\'ere/Greisen radius used in the LDF functions may lead to determine the shape of the $s_{\rm{local}}(r)$ or $\beta_{\rm{ss}}(r)$ versus $r$ curves. The frequency distributions of the minimum values of $s_{\rm{local}}(r)$ for two different Moli\'ere radii possess different mean values. This suggests that the scale distance regulates the shape of the $s_{\rm{local}}(r)$ or $\beta_{\rm{ss}}(r)$ versus $r$ curves on a statistical basis.

4. The KASCADE experimental data in figures 4, 5 and 6 in terms of the important measured parameters (mean minimum LAP and mean maximum SSP) indicate that the CR mass composition follows a gradual change from predominantly lighter (proton-dominated) to heavier (iron-dominated) nuclei with the increase of shower size and muon size or energy. This above feature supports the results obtained from the study of the $N_{\rm e}\textendash N_{\mu}$ variations in the KASCADE experiment.      

\section*{Acknowledgment}
The authors sincerely thank Dr. J. Wochele, KIT, Germany for useful suggestions on some technical issues about KCDC. The authors would like to thank both the reviewers for their detailed and useful comments. RKD acknowledges the financial support from the SERB, Department of Science and Technology (Govt. of India) under the Grant no. EMR/2015/001390.

\end{document}